\documentclass[12pt]{elsart4-1}

\usepackage{amssymb}
\usepackage{graphicx}
\usepackage{amssymb}
\usepackage{amsfonts}
\usepackage{amsbsy}

\DeclareGraphicsExtensions{.eps}

\def\FIGURE#1#2#3#4{
\begin{figure}
  \centering 
  \includegraphics[#2]{#1}
  \caption{#4} 
  #3
\end{figure}
}

\newcommand\cal{\mathcal}

\usepackage[english,francais]{babel}

\newtheorem{e-proposition}[theorem]{Proposition}

\newtheorem{e-definition}[theorem]{Definition\rm}

\renewcommand{\theequation}{\arabic{equation}}
\def\og{\leavevmode\raise.3ex\hbox{$\scriptscriptstyle\langle\!\langle$~}}
\def\fg{\leavevmode\raise.3ex\hbox{~$\!\scriptscriptstyle\,\rangle\!\rangle$}}

\def\IIb{{IIb}}
\def\blue{}
\newfam\black
\font\blackboard=msbm10 
\font\blackboards=msbm7
\font\blackboardss=msbm5
\textfont\black=\blackboard
\scriptfont\black=\blackboards
\scriptscriptfont\black=\blackboardss
\def\Bbb#1{{\fam\black\relax#1}}

\def\CC{{\cal C}}

\def\CF{{\cal F}}

\def\CM{{\cal M}}
\def\CN{{\cal N}}

\def\BZ{\Bbb{Z}}

\def\GeV{~{\rm GeV}}

\def\Im{{\rm Im}~}

\def\pp{\partial}

\def\II{{II}}
\def\IIa{{IIa}}

\begin{document}

\begin{frontmatter}


%
\selectlanguage{english}
\title{Basic results in Vacuum Statistics}

\selectlanguage{english}
\author[1,2]{Michael R. Douglas}
\ead{mrd@physics.rutgers.edu}
\address[1]{NHETC and Department of Physics and Astronomy\\
Rutgers University\\
Piscataway, NJ 08855--0849, USA}
\address[2]{I.H.E.S., Le Bois-Marie, Bures-sur-Yvette, 91440 France}

\begin{abstract}

A review of recent work on constructing and finding
statistics of string theory vacua,
done in collaboration with Frederik Denef, Bogdan Florea,
Bernard Shiffman and Steve Zelditch.

To appear in the proceedings of Strings 2004, June 28-July 2, Paris, France.

\end{abstract}
\end{frontmatter}



\selectlanguage{english}

\begin{quotation}
Tout ce qui est simple est faux, 
mais tout ce qui ne l'est pas est inutilisable. 

-- Paul Val\'ery
\end{quotation}

\section{Predictions from string theory}

For almost 20 years we have had good qualitative arguments that
compactification of string theory can reproduce the Standard Model
and solve its problems, such as the hierarchy problem.  But we
still seek distinctive predictions which 
we could regard as evidence for or against the theory.

One early spin-off of string theory, four dimensional supersymmetry,
is the foundation of most current thinking in ``beyond the Standard
Model'' physics.  Low energy supersymmetry appears to fit well with
string compactification.  But would not discovering
supersymmetry be evidence against string/M theory?

In recent years, even more dramatic possibilities have been suggested,
which would lead to new, distinctive particles or phenomena:
large extra dimensions (KK modes);
a low fundamental string scale (massive string modes);
rapidly varying warp factors (modes bound to branes;
conformal subsectors), and so on.
Any of these could lead to dramatic discoveries.  But should we expect
string/M theory to lead to any of these possibilities?  
Would not discovering them be evidence against string theory?

At Strings 2003, I discussed a {\blue statistical} approach to these
and other questions of string phenomenology.  Over the last year, our
group at Rutgers, and the Stanford group, have made major progress in
developing this approach, with
\begin{itemize}
\item Explicit proposals for vacua with all moduli stabilized along the
lines of {\blue KKLT} \cite{KKLT} (work with Denef and Florea \cite{DDF}).
\item Detailed results for distributions of these vacua (with Shiffman,
Zelditch and Denef \cite{DSZ,DD})
\item Preliminary results on the statistics of 
the volume of the extra dimensions (in progress with Denef),
and on supersymmetry breaking scales 
\cite{mrd-susy,toapp,Suss-susy}.
\end{itemize}

These ideas have already begun to inspire new phenomenological models
(for example \cite{AHD,Giudice:2004tc}).  
Even better, if we are lucky and the number of
vacua is not too large (as we explain below),
fairly convincing predictions might come out of this
approach over the next few years.
While much work would be needed to bring this about,
we may be close to making
some predictions: those which use just the most generic features of
string/M theory compactification, namely the existence of many hidden sectors.

\section{Hidden sectors}

Before string theory, and during the ``first superstring revolution,''
most thinking on unified theories assumed that internal consistency
of the theory
would single out the matter content we see in the real world.
In the early 1980's it was thought that
$d=11$ supergravity might do this.  
Much of the early excitement about
the heterotic string came from the fact that it could easily produce
the matter content of $E_6$, $SO(10)$ or $SU(5)$ grand unified
theory.

But this was not looking at the whole theory.  The typical
compactification of heterotic or type \II\ strings on a Calabi-Yau
manifold has hundreds of scalar fields, larger gauge groups and more
charged matter.  Already in the perturbative heterotic string an extra
$E_8$ appeared.  With branes and non-perturbative gauge symmetry, far
larger groups are possible, with many simple factors.
If we live in a ``typical'' string compactification, it seems
that there are many hidden sectors, not directly visible to
observation or experiment.

Should we care?  Does this lead to any general predictions?
Hidden sectors may or may not lead to new particles or forces.  But
what they do generically lead to is a {multiplicity of vacua},
because of symmetry breaking, choice of vev of additional scalar
fields, or other discrete choices.

Let us say a hidden sector allows $c$ distinct vacua or ``phases.''
If there are $N$ hidden sectors, the multiplicity of vacua will go as
$$
{\cal N}_{vac} \sim c^N .
$$

While the many hidden sectors certainly make the detailed study of
string compactification more complicated, we should consider the idea
that they lead to simplifications as well.
Thus we might ask, what can we say about the case of a large number $N$
of hidden sectors ?  Clearly there will be a {\blue large} multiplicity
of vacua.

We only live in one vacuum.  However, as pointed out by {\blue Brown
and Teitelboim \cite{BT}, Banks, Dine and Seiberg \cite{BDS}}, and no
doubt many others, vacuum multiplicity can help in solving the
cosmological constant (c.c.) problem.  In an ensemble of $\CN_{vac}$ vacua
with roughly uniformly distributed c.c. $\Lambda$, one expects that
vacua will exist with $\Lambda$ as small as $M_{pl}^4/\CN_{vac}$.
To obtain the observed small nonzero c.c.  $\Lambda \sim 10^{-122}
M_{pl}^4$, one requires $\CN_{vac} > 10^{120}$ or so.

Now, assuming different phases have different vacuum energies, adding
the energies from different hidden sectors can produce roughly uniform
distributions.  In fact, the necessary $\CN_{vac}$ can easily be fit
with $\CN_{vac} \sim c^N$ and
the parameters $c\sim 10$, $N\sim 100-500$ one expects from
flux compactification of string theory, 
as first argued by {\blue Bousso and
Polchinski} \cite{BouPol}.

One might regard fitting the observed small nonzero c.c.
in {\bf any} otherwise acceptable vacuum
as solving the problem, or one might appeal to an anthropic argument
such as that of {\blue Weinberg} \cite{Weinberg} to select this vacuum.  
In the absence of other candidate solutions to the problem, we might even
turn this around and call these ideas 
{\bf evidence} for the hypothesis that
we are in a compactification with many hidden sectors.

\section{Supersymmetry breaking}

So can we go further with these ideas?  Another quantity which can get
additive contributions from different sectors is the {\blue scale of
supersymmetry breaking}.  Let us call this $M_{susy}^2$ (we will define
it more carefully below).

We recall the classic arguments for low energy supersymmetry from
{\blue naturalness}.  The electroweak scale $m_{EW}$ is far below the
other scales in nature $M_{pl}$ and $M_{GUT}$.  According to one
definition of naturalness, this is only to be expected if a symmetry
is restored in the limit $m_{EW}\rightarrow 0$.  This is not true if
$m_{EW}$ is controlled by a scalar (Higgs) mass $m_H$, but can be true
if the Higgs has a supersymmetric partner (we then restore a chiral
symmetry).

A more general definition of naturalness requires the theory to be
stable under radiative corrections, so that the small quantity does not
require fine tuning.  Again, low energy supersymmetry can accomplish this.
Many theories have been constructed in which 
$$
M_H^2 \sim c M_{susy}^2 ,
$$
with $c \sim 1/10$ without fine tuning.  Present data typically requires
$c < 1/100$, which requires a small fine tuning (the ``little hierarchy
problem.'')

On the other hand, the solution to the cosmological constant problem we
accepted above, in terms of a discretuum of vacua,
is suspiciously similar to fine tuning the c.c., putting the role
of naturalness in doubt.  What should replace it?

The original intuition of string theorists was that string theory would
lead to a {\blue unique} four dimensional vacuum state, or at most very
few, such that only one would be a candidate to describe real world physics.
In this situation, there is no clear reason the unique theory should be
``natural'' in the previously understood sense.

With the development of string compactification, it has become
increasingly clear that there is a large multiplicity of vacua.
The vacua differ not only in the cosmological constant, but in every
possible way: gauge group, matter content, couplings, etc.
What should we do in this situation?

The ``obvious'' thing to do at present is to make the following definition
\cite{mrd-stat}:
\begin{quotation}
An effective field theory (or specific coupling, or observable) $T_1$ is
{\bf more natural} in string theory than $T_2$, if the {\blue number} of 
phenomenologically acceptable vacua
leading to $T_1$ is larger than the {\blue number} leading to $T_2$.
\end{quotation}

Now there is some ambiguity in defining ``phenomenologically
acceptable'' (or even
``anthropically acceptable,'' as some would have it \cite{Suss-anth}).  
One clearly wants $d=4$, supersymmetry
breaking, etc.  One may or may not want to put in more detailed information
from the start.

In any case, the unambiguously defined information
provided by string/M theory is the {\blue number of vacua} and the
{\blue distribution of resulting EFT's}.
For example, we could define
\begin{eqnarray*} \label{eq:susydist}
d\mu[M_{H}^2,M_{susy}^2,\Lambda] =
 \rho(M_{H}^2,M_{susy}^2,\Lambda) dM_{H}^2\ dM_{susy}^2\ d\Lambda \hfill\\
 = \sum_{T_i} \delta(M_{susy}^2 - M_{susy}^2|_{T_i})
 \delta(M_{H}^2 - M_{H}^2|_{T_i})
 \delta(\Lambda-\Lambda|_{T_i})
\end{eqnarray*}
a distribution which counts vacua with given c.c., susy breaking scale and
Higgs mass, and study the function
$$
\rho(10^4 \GeV^2, M_{susy}^2, \Lambda \sim 0) .
$$

\section{Statistical selection}

Is this definition of ``stringy naturalness'' good for anything?
Suppose property $X$ (say low scale susy) were realized by $10^{40}$
phenomenologically acceptable
vacua, while $\bar X$ (say high scale susy) were realized by $10^{20}$ 
such vacua.
If by prediction we mean not just a hunch or a reason to bet on a
particular property, but a property whose observation would actually
falsify string theory (and this is what we really need in the end),
we should {\bf not} conclude that string theory predicts low scale
susy.  

On the other hand, if the distribution is sharply enough peaked, and there
are not too many vacua, it could well turn out that some regions of
theory space would have {\bf no} vacua, and we would get a prediction.

For example, suppose there were $10^{160}$ vacua 
with the property $X$ (say low scale susy), and which
realize all known physics, except for the observed c.c..
Suppose further that they realize
a uniform distribution of cosmological constants; then out of this
set we would expect about $10^{40}$ to also reproduce the observed
cosmological constant.  Suppose furthermore that $10^{100}$ vacua with
property $\bar X$ work except for possibly the c.c.; out of this set
we only expect the correct c.c. to come out if an additional
$10^{-20}$ fine tuning is present in one of the vacua which comes
close.  Not having any reason to expect this, and having other vacua
which work, we have reasonable grounds for predicting $X$, in the
strong sense that observing $\bar X$ would be evidence {\bf against}
string theory.

In a systematic approach, one would take all aspects of the physics
resulting from each choice of vacuum, not just the c.c. but couplings
and matter content as well, and make the analogous argument.  As
discussed in \cite{mrd-stat}, the rest of the information at hand is
comparable in selectivity to the c.c.; say a rough fraction
$10^{-240}$ of vacua out of a fairly uniform ensemble might reproduce
the Standard Model, and thus this is an important improvement.
However the basic idea leading to predictions is more or less the same.

Upon considering the entire problem in this way, the most crucial
advantage of the statistical approach becomes apparent.  It is that we
can benefit by the hypothesis that some properties of the distribution
of vacua are (to a good approximation) statistically independent, in
which case we can argue that vacua exist which realize a group of properties
simultaneously, even without finding explicit examples.

For example, it seems very likely that the value
of the c.c. is independent of the number of Standard Model
generations, in the sense that even if we restrict attention to vacua
with a given number $N_{gen}$ of generations we will still find a
uniform distribution of c.c.'s with cutoff independent of $N_{gen}$.
Then, suppose for sake of argument that a fraction $10^{-4}$ of the
vacua have $N_{gen}=3$.  While not rare compared to other properties,
this is sufficiently rare to make it significantly more difficult to
find models with both $N_{gen}=3$ and the small c.c..  Rather than do
this, we should study the larger population of models with arbitrary
$N_{gen}$ and check the hypothesis that these properties are
independent.  Having done this, we can argue that the fraction of
vacua which realize both properties is the product of the fractions
which realize each, without explicitly finding the vacua which realize
both.  Of course, the independence hypothesis might turn out to be false;
if we found evidence of a correlation between
$N_{gen}$ and the c.c., that would be even more interesting (and
surprising).  

We can go on to make the same type of analysis for
each of the characteristic
properties of the Standard Model
(the gauge group, the hierarchy, the details of the couplings
and so on).  While any one of its specific properties
is ``rare'' in the sense that the great majority of
vacua do not realize it (most vacua will not have unbroken gauge
symmetry at low energy, etc.), it seems unlikely that any one of them
(even the c.c.) is so rare as to allow only a few candidate vacua.
Multiplying the fractions of vacua which realize the various properties
leads to an estimated fraction 
of vacua which agree with the Standard Model, finite
but so small that the task of finding the vacuum which actually
realizes all of its properties simultaneously is almost impossible.
On the other hand, by separating the various properties of interest
into subsets, such that correlations are possible only within each subset,
we can hope to divide up the problem into manageable pieces.

These arguments and examples illustrate how, under certain possible
outcomes for the actual number and distribution of vacua, we could 
make well motivated predictions.  Of course the actual
numbers and distribution are not up to us to chose, and one can
equally well imagine scenarios in which this type of
predictivity is not possible.  For example, $\CN_{vac} \sim 10^{1000}$
would probably not lead to predictions, unless the distribution were
very sharply peaked, or unless we make further assumptions which
drastically cut down the number of vacua.

\section{Absolute numbers}

The basic estimate for numbers of flux vacua \cite{AD} is 
$$
{\cal N}_{vac} \sim \frac{(2\pi L)^{K/2}}{(K/2)!} [c_n]
$$
where $K$ is the number of
distinct fluxes ($K=2b_3$ for \IIb\ on CY$_3$) and $L$ is a ``tadpole
charge'' ($L=\chi/24$ in terms of the related CY$_4$).  The ``geometric
factor'' $[c_n]$ does not change this much, while other multiplicities are
probably subdominant to this one.

Typical $K\sim 100-400$ and $L\sim 500-5000$, leading to 
${\cal N}_{vac} \sim 10^{500}$.  This is probably too large for 
statistical selection to work.

On the other hand, this estimate did not put in all the consistency
conditions.  Here are two ideas, still rather speculative.
\begin{itemize}
\item
Perhaps stabilizing the moduli not yet considered in detail (e.g. brane
moduli) is highly non-generic, or perhaps
most of the flux vacua become unstable after supersymmetry
breaking due to KK or stringy
modes becoming tachyonic.  At present there is no evidence for these ideas,
but neither have they been ruled out.

\item
Perhaps cosmological selection is important: almost all vacua have
negligible probability to come from the ``preferred initial
conditions.''  Negligible means $P <<< 1/\CN_{vac}$, and almost all
existing proposals
for wave functions or probabilty factors are not so highly peaked, but
{\bf eternal inflation} has been claimed to be
(as reviewed in \cite{Guth}),
and it is important to know if this is relevant for string theory 
(see for example \cite{FreiSuss}).
\end{itemize}

Such considerations might drastically cut the number of vacua.  While
we would then need to incorporate these effects in the distribution,
it is conceivable that to a good approximation these effects are
{\blue statistically independent} of the properties of the distribution
which concern us, so that the statistics we are computing now are the
relevant ones.  Even if not, it seems very unlikely to us that 
cosmology will select a unique vacuum {\it a priori}; rather
we believe the problem with these considerations taken
into account will not look so different formally (and perhaps even
physically) from the problem without them, and thus we proceed.

\section{Stringy naturalness}

The upshot of the previous discussion is that in this picture,
either string theory is not predictive because there are too many vacua, 
or else the key to making predictions is to count vacua, find their
distributions, and apply the principles of statistical selection.

To summarize this, we again oversimplify and describe statistical selection
as follows: we propose to show that a property $\bar X$ cannot come
out of string theory by arguing that {\bf no} vacuum realizing $\bar
X$ reproduces the observed small c.c. (actually, we are considering
all properties along with the c.c.).  One might ask how we can hope
to do this, given that computing the c.c. in a specific vacuum to the
required accuracy is far beyond our abilities.  The point is that it
should be far easier to characterize the distribution of c.c.'s than
to compute the c.c. in any specific vacuum.  
To illustrate, suppose we can compute
it at tree level, but that these results receive complicated perturbative and
non-perturbative corrections.  Rather than compute these exactly in
each vacuum, we could try to show that they are uncorrelated with
the tree level c.c.; if true and if the tree level distribution is
simple (say uniform), the final distribution will also be simple.

If so, tractable approximations to the true distribution of vacua
can estimate {\blue how
much unexplained fine tuning} is required to achieve the desired EFT,
and this is the underlying significance of the definition of ``stringy 
naturalness'' we gave above.

Thus, we need to establish that vacua satisfying the various requirements
exist, and estimate their distribution.  We now discuss results on these
two problems, and finally return to the question of the distribution of
supersymmetry breaking scales.

\section{Constructing KKLT vacua}

The problem of stabilizing all moduli in a concrete way in string
compactification has been studied for almost 20 years.  One of the
early approaches 
was to derive an effective Lagrangian by KK reduction,
find a limit in which nonperturbative effects are small, and add 
sufficiently many nonperturbative corrections to produce a generic
effective potential.  Such a generic potential, depending on all
moduli, will have isolated minima.
While the idea is simple, the complexities of string compactification
and the presence of hundreds of moduli have made it hard to carry out.

A big step forward was the development of flux compactification by
Polchinski and Strominger \cite{Polchinski:1995sm};
Becker and Becker \cite{Becker:1996gj};
Dasgupta, Rajesh and Sethi \cite{Dasgupta:1999ss},
and many others.
Since the energy of fluxes in the compactification manifold
depends on moduli, turning on flux allows stabilizing a large subset of
moduli at the classical level.
{\blue Acharya} \cite{Acharya}
has proposed that in $G_2$ compactification, all metric
moduli could be stabilized by fluxes.  However it is not yet known how
to make explicit computations in this framework.

The most computable class of flux compactifications at present is that
of {\blue Giddings, Kachru and Polchinski} \cite{GKP}, in \IIb\ orientifold
compactification, because one can appeal to the highly developed theory
of Calabi-Yau moduli spaces and periods.
However the \IIb\ flux superpotential does not depend on K\"ahler
moduli, nor does it depend on brane or bundle moduli.  Now one can
argue that the brane/bundle moduli parameterize {\blue compact} moduli
spaces ({\it e.g.} consider the D3 brane), and thus they will be
stabilized by a {\blue generic} effective potential.
However, for the K\"ahler moduli, we need to show that the minimum is
not at infinite volume, or deep in the stringy regime (in which case
we lose control).  Thus we need a fairly explicit expression for their
effective potential.

{\blue Kachru, Kallosh, Linde and Trivedi (KKLT) \cite{KKLT}
proposed to combine the flux superpotential with
\begin{itemize}
\item a nonperturbative superpotential produced by D3 instantons
and/or D7 world-volume gauge theory effects.  These depend on K\"ahler
moduli and can in principle fix them.
\item energy from an anti-D3 brane, which would break supersymmetry
and lift the c.c. to a small positive value.
\end{itemize}
However they did not propose a concrete model which contained these 
effects, and in fact such models are not so simple to find.
The main problem is that most brane gauge theories in
compactifications which cancel tadpoles, have too much matter to
generate superpotentials.  One needs systematic techniques to determine
this matter content, or compute instanton corrections, and find
the examples which work.

In \cite{DDF}, Denef, Florea and I found the first 
examples which 
work.  Our construction relies heavily on the analysis of
instanton corrections in F theory due to {\blue Witten \cite{Witten}, 
Donagi} and
especially {\blue A. Grassi \cite{Grassi}.
Their starting point was to compactify M theory on a Calabi-Yau
fourfold $X$.  This leads to a 3D theory with four supercharges, 
related to F theory and \IIb\ if $X$ is $T^2$-fibered,
by taking the limit ${\rm vol}(T^2)\rightarrow 0$.  The complex
modulus of the $T^2$ becomes the dilaton-axion varying on the base $B$.

In M theory, an {\blue M5 brane}  wrapped on a
divisor $D$ (essentially, a hypersurface),
will produce a nonperturbative superpotential, if $D$ has
{\blue arithmetic genus one}:
$$
1 = \chi({\cal O}_D) = h^{0,0} - h^{0,1} + h^{0,2} - h^{0,3} .
$$
Each of these complex cohomology groups leads to two
fermion zero modes; an instanton contributes to $W$ if there are 
exactly two.  
A known subset of these (vertical divisors) survive in F theory.

In the F theory limit, only divisors which wrap the $T^2$ contribute,
and these correspond to D3-instantons wrapping surfaces in $B$.
Thus, we looked for F theory
compactifications on an elliptically fibered fourfold $X$ with enough
divisors of arithmetic genus one, so that a superpotential
$$
W = W_{flux} + \sum_i b_i e^{-\vec t\cdot D_i}
$$
will lead to non-trivial solutions to $DW=0$ by balancing 
the exponentials against the dependence coming from the K\"ahler 
potential.

In the math literature, there is a very general relation between
divisors of a.g. one, and contractions of manifolds.  It implies that
{\blue no model with one K\"ahler modulus} can stabilize K\"ahler
moduli (see also \cite{RobbinsSethi}), at least by
using a.g. one divisors.  
Now there may be ways beyond the a.g. one condition: {\blue Witten} 
\cite{Witten} suggested
that $\chi(D)>1$ might work as well (without providing examples);
furthermore Gorlich {\it et al}  \cite{KTflux} have argued that
{\blue flux} lifts additional matter
and relaxes some of these constraints.

In any case, there is no problem if the CY threefold has more than one
K\"ahler modulus, as in the vast majority of cases.
Using the very complete study of divisors of a. g. one of
A. Grassi \cite{Grassi}, we have found $6$ models 
with toric Fano threefold base which can stabilize all K\"ahler moduli,
and could be analyzed in detail using existing techniques.

The simplest, ${\cal F}_{18}$, has $89$ complex structure moduli.
According to the AD counting formula, it should have roughly
$\epsilon\times 10^{307}$ flux vacua with all moduli stabilized,
where 
$$
\epsilon = g_s \times \frac{|W|^2}{m_s^6} \bigg|_{max} .
$$

Models which stabilize all K\"ahler moduli using a.g. one divisors
are {\blue not} generic, but they are not uncommon
either; there are 29 out of 92 with Fano base, and probably many
more with ${\mathbb P}^1$ fibered base.  This last class of model should be
simpler, in part because they have heterotic duals, but analyzing
them requires better working out the D7 world-volume theories.
We also expect one can add antibranes or D breaking as in the KKLT discussion
to get de Sitter vacua, but have not yet analyzed this.

\section{Flux vacua}

\FIGURE{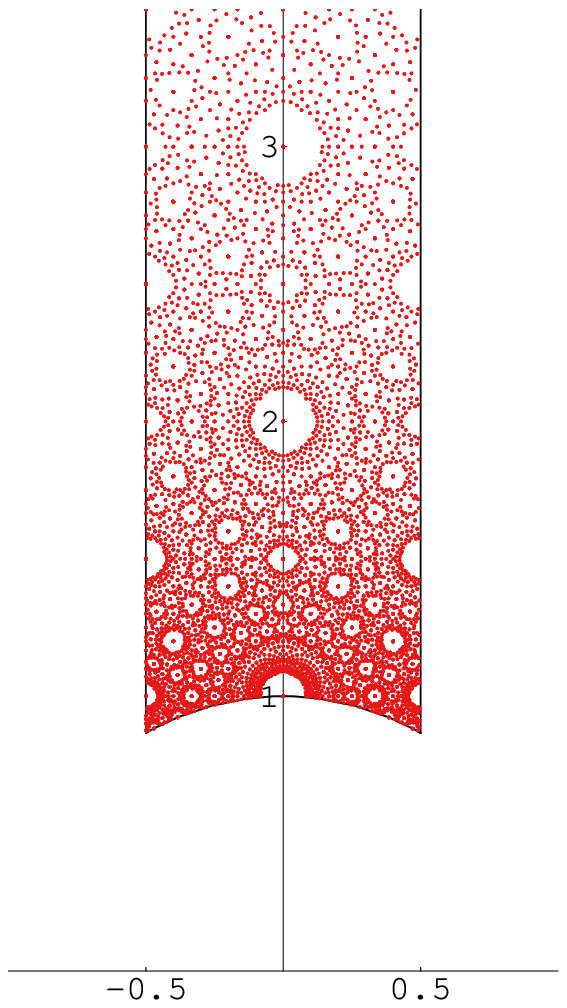}{height=9.5cm}{\label{T2vacua}}%
{Values of $\tau$ for rigid CY flux vacua with $L_{max} = 150$,%
 from \cite{DD}.}

We recall the ``flux superpotential'' of {\blue Gukov,
Taylor, Vafa and Witten}
in \IIb\ string on CY,
$$
W = \int \Omega(z) \wedge \left( F^{(3)} + \tau H^{(3)} \right).
$$
The simplest example
is to consider a {\blue rigid} CY, {\it i.e.} with $b^{2,1}=0$
(for example, the orbifold $T^6/\BZ_3$).
Then the only modulus is the dilaton $\tau$, with K\"ahler potential
$K=-\log\Im\tau$, and the flux superpotential
reduces to
$$
W = A \tau + B; \qquad A=a_1 + \Pi a_2; B=b_1 + \Pi b_2
$$
with $\Pi=\int_{\Sigma_2} \Omega^{(3)}/\int_{\Sigma_1}\Omega^{(3)}$,
a constant determined by CY geometry.

Now it is easy to solve the equation $DW=0$:
\begin{eqnarray*}
DW &= \frac{\pp W}{\pp\tau} - \frac{1}{\tau-\bar\tau}W 
 &= \frac{-A\bar\tau -B}{\tau-\bar\tau}
\end{eqnarray*}
so $DW=0$ at
$\bar\tau = -\frac{B}{A}$
where $\bar\tau$ is the complex conjugate.

The resulting set of flux vacua for $L=150$ and $\Pi=i$ is shown in
Fig. \ref{T2vacua}.
A similar enumeration for a Calabi-Yau with $n$ complex structure moduli,
would produce a similar plot in $n+1$ complex dimensions, 
the distribution of flux vacua.  It could (in principle) be mapped into
the distribution of {\blue possible values of coupling constants}
in a physical theory.

This intricate distribution has some simple properties.
For example, one can get exact results for the large $L$ asymptotics,
by computing a continuous distribution $\rho(z,\tau;L)$, whose integral
over a region $R$ in moduli space reproduces the  asymptotic number of
vacua which stabilize moduli in the region $R$, for large $L$,
$$
\int_R dz d\tau\ \rho(z,\tau;\;L) \sim_{L\rightarrow\infty} N(R) .
$$

\FIGURE{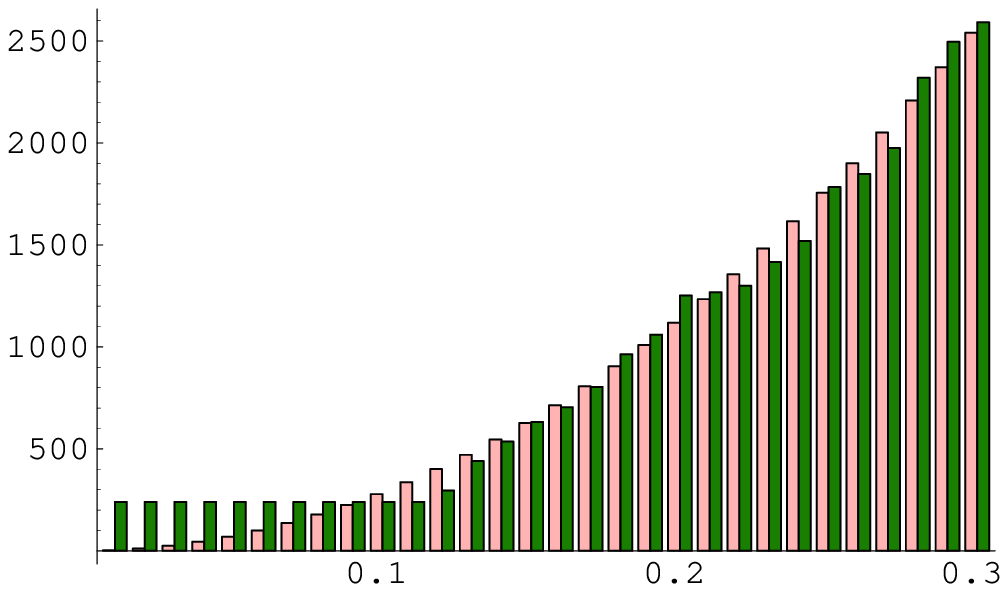}{width=3in}{  \label{discs}}%
{Number of
vacua in a circle of coordinate radius $R$ around $\tau = 2i$, with
$R$ increasing in steps $dR = 0.01$. Pink bars give the estimated
value, green bars the actual value.}

For a region of radius $r$, the continuous approximation should become
good for $L >> K/r^2$.  For example, if we consider a circle of radius
$r$ around $\tau=2i$, we match on to the constant density distribution
for $r > \sqrt{K/L}$, as we see in Fig. \ref{discs}.

Explicit formulae for the continuous densities can be found, in terms of the
geometry of the moduli space $\CC$.
The simplest such result \cite{AD}
computes the {\blue index density} of vacua:
$$
\rho_I(z,\tau) = \frac{(2\pi L)^{b_3}}{b_3!\pi^{n+1}} \det(-R-\omega\cdot 1)
$$
where $\omega$ is the K\"ahler form and $R$ is the matrix of curvature
two-forms.  
Integrating this over a fundamental region of the moduli space
produces an estimate for
the total number of flux vacua.  For example, for $T^6$ we
found $I \sim 4\cdot 10^{21}$ for $L=32$.
Since $r \sim 1$ in the bulk of moduli space, the condition $L >
K/r^2$ for the validity of this estimate should be satisfied.

Another example is the ``mirror quintic,''
with a one parameter moduli space $\CM_c(\tilde Q)$ \cite{DD}.
The integral is
$$
\frac{1}{\pi^2}\int_\CC \det(-R-\omega) = 
\frac{1}{12}\chi(\CM_c(\tilde Q)) = \frac{1}{60} .
$$

This density is ``topological'' and there are mathematical techniques
for integrating it over general CY moduli spaces \cite{Lu-MRD}.
Good estimates for the index should become
available for a large class of CY's over the coming years.

\section{Distributions of flux vacua}

\FIGURE{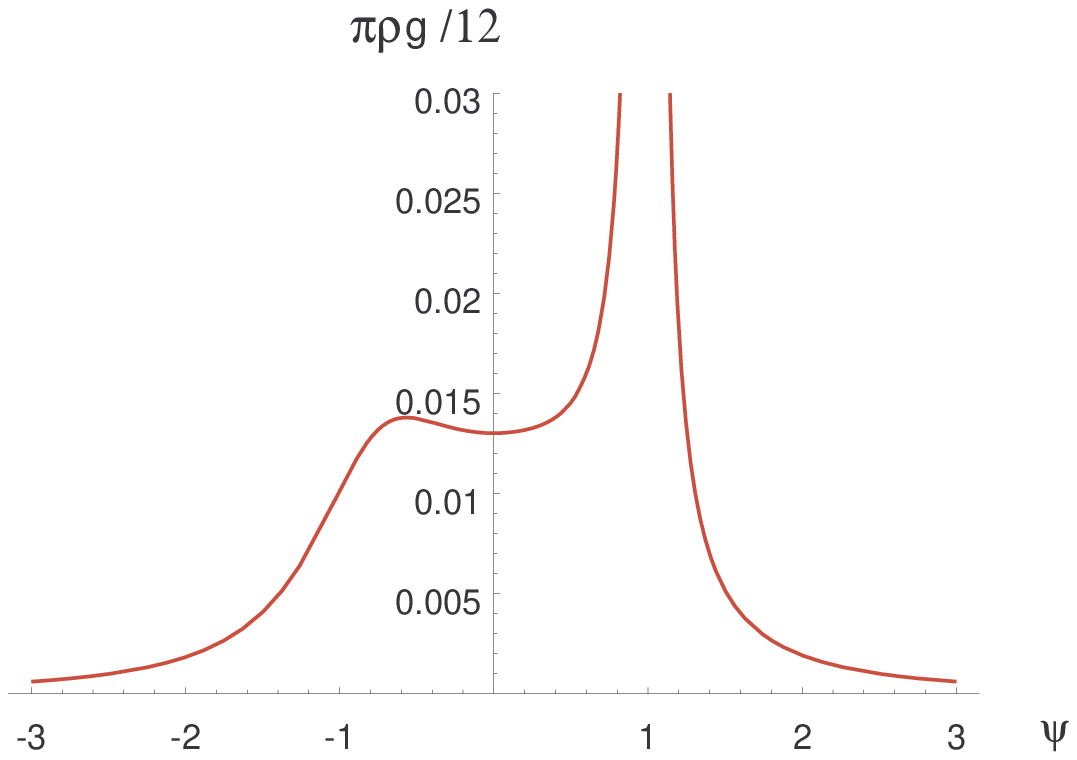}{width=3in}{  \label{quinticsusydens}}%
{The susy vacuum number density per unit
$\psi$ coordinate volume, $\pi \rho g_{\psi \bar{\psi}}/12$, on
the real $\psi$-axis, for the mirror quintic.}

In Fig. \ref{quinticsusydens}, we see the detailed distribution
of flux vacua on the mirror quintic ($K=4$ and $n=1$), on a real slice
through the complex structure moduli space \cite{DD}.
Note the divergence at $\psi=1$.  This is the conifold point, with
a dual gauge theory interpretation.  It arises because the curvature
$R \sim \pp{\bar\pp}\log\log |\psi-1|^2$ diverges there.
The divergence is integrable,
but a finite fraction of all the flux vacua sit near it.

Quantitatively, for the mirror quintic,
\begin{itemize}
\item About $3\%$ of vacua sit near the conifold point, with an
induced scale $|\psi-1| < 10^{-3}$.
\item About $1\%$ of vacua have $|\psi-1| < 10^{-10}$.  More generally,
the density and number of vacua with $S\equiv\psi-1$ goes as
$$
\rho_{vac} \sim \frac{d^2S}{|S\log S|^2}; \qquad
\CN_{vac}|_{S<S_*} \sim \frac{1}{|\log S_*|} .
$$
Writing $S=e^{-1/g^2}$, this is $\rho \sim d^2g/|g|^2$.
\item About $36\%$ of vacua are in the ``large complex structure limit,''
defined as $\Im t>2$ with $5\psi=e^{2\pi it/5}$.
Here $\rho \sim d^2t/(\Im t)^2$.
\end{itemize}

Vacua close to conifold degenerations are interesting for model
building, as they provide a natural mechanism for generating large
hierarchies.  We have found that such vacua are common, but are by no
means the majority of vacua.  

Note that some of these vacua have a dual
gauge theory interpretation, while in other cases they should be
thought of as supergravity backgrounds, possibly leading to
Randall-Sundrum phenomenology \cite{RandallSundrum}.  
The question of which interpretation is appropriate depends on the
local parameter $g_s N$ where $N$ is the specific flux dual to the
number of branes, and on the embedding of the Standard Model.  It
would be very interesting of course if one or the other alternative
were strongly favored in the vacuum counting.  Assuming the uniform
distributions $d g_s$ and $dN$, one might expect a preference for $g_s
N>1$ and Randall-Sundrum, but the ratio would seem to be at most of order
$10^2-10^3$ which could easily be outweighed by other considerations.
To do this right, one should also factor in the (model-dependent)
relation between $g_s$ and the known (string scale) Standard Model
couplings $g \sim 1/25$, which should lower the expected $g_s N$
and further weaken this preference.

\FIGURE{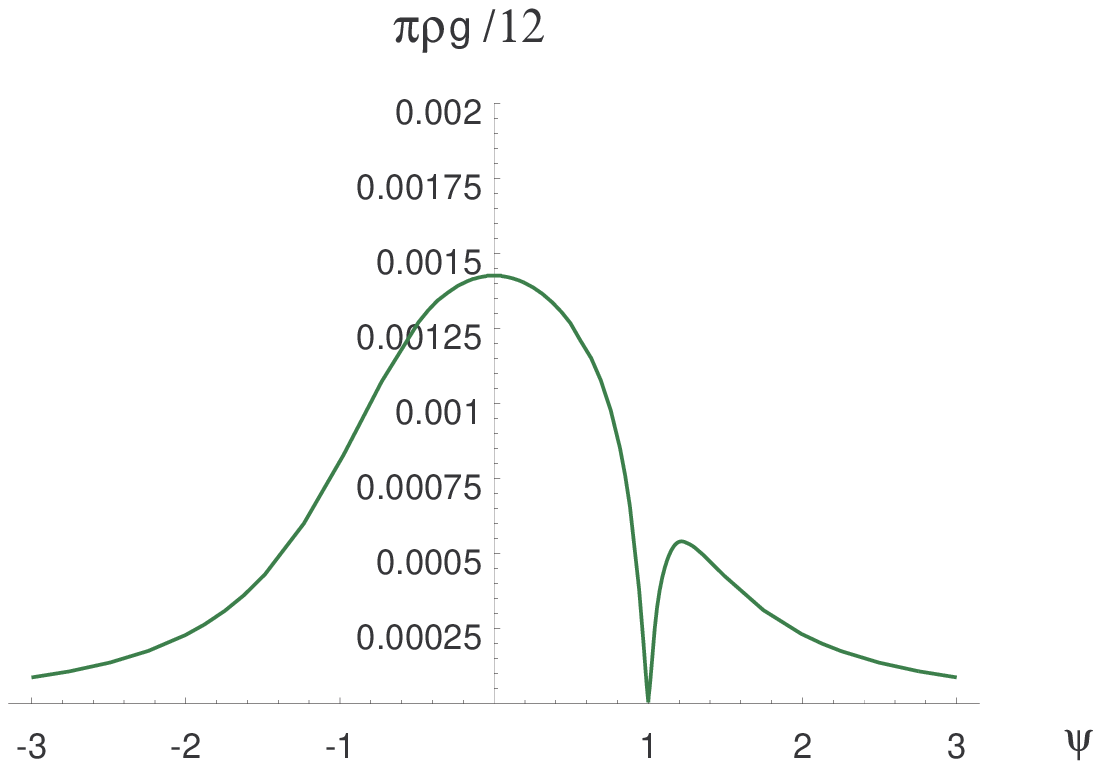}{width=3in}%
{\label{quinticsusyposMdens}}%
{The number density of susy vacua with positive mass matrix,
per unit coordinate volume, on the real
$\psi$-axis, for the mirror quintic. Compare to fig. %
\ref{quinticsusydens}.}

Suppose we go on to break supersymmetry by adding an anti D3-brane, or
by other D term effects.  The previous analysis applies (since we have
not changed the F terms), but now it is necessary that the mass matrix
at the critical point is positive.
The resulting distribution of tachyon-free D breaking vacua is given
in Fig. \ref{quinticsusyposMdens}.
In fact, most D breaking vacua near the conifold point have tachyons
(for one modulus CY's), so we get suppression, not enhancement.
This is not hard to understand in detail; the mechanism is a sort of
``seesaw'' mixing between modulus and dilaton, which seems special
to one parameter models.

\begin{figure}[t]
  \centering 
  \includegraphics[height=5.1cm]{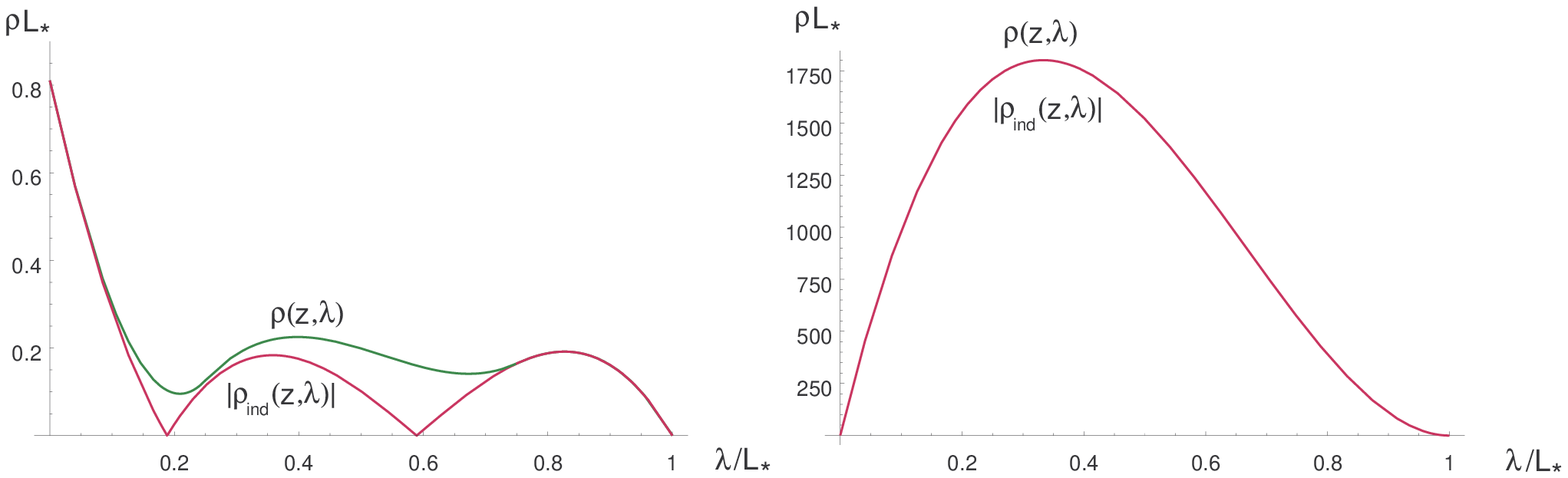}
  \caption{\emph{Left:} Vacuum number densities (true and absolute value of
index) in the large complex structure limit ($\CF=2/\sqrt{3}$), as
a function of cosmological constant value,
from \cite{DD}. \emph{Right:} Same near
the conifold limit ($\CF=100$ for this example).}
  \label{CCdistr}
\end{figure}

In Fig. \ref{CCdistr} one sees
the distribution of (negative) AdS cosmological constants
$\hat \Lambda=3e^K|W|^2$, both at generic
points (left) and near the conifold point (right).  Note that at
generic points it is fairly uniform, all the way to the string scale.
On the other hand, imposing small c.c.  competes with 
the enhancement of vacua near the conifold point.
The left hand graph compares the total number of vacua (green) with
the index (red).  The difference measures the number of 
{\blue K\"ahler stabilized vacua}, vacua which exist because of the
structure of the K\"ahler potential, not the superpotential.

\section{Large complex structure/volume}

Another simple universal property: given $n>>1$ moduli,
the number of vacua falls off
rapidly at large complex structure, or in a \IIa\ mirror picture
at large volume $V$, as
$$
\int_{V>V_0} \rho \sim V_0^{-n/3} .
$$

To see this, first note that $R\sim\omega$ in this regime, so the
distribution of vacua is determined by the volume form derived
from the {\blue metric on the space of metrics},
$$
\left< \delta g_{ij}, \delta g_{kl} \right> =
{1\over V}\int_{CY} \sqrt{g} g^{ik} g^{jl}
\delta g_{ij} \delta g_{kl}
$$
The $1/V$ factor 
(which compensates the $\sqrt{g}$) comes from the standard derivation
of the kinetic term in KK reduction on CY.
Because of the inverse
factors of the metric, this falls off with volume as $V^{-1/3}$.
Since the volume form is $\sqrt{G}$,
this factor appears for {\blue each} modulus.
For large $n$, $\CN\sim V^{-n/3}$
is a drastic falloff, and (as we saw explicitly in an example in 
\cite{DDF}) typically there are no vacua in this regime.

A possible physical application of this: we know how to stabilize
complex structure moduli using fluxes in \IIb.  
Suppose we can use T-duality to get a corresponding class of models
in \IIa\ with stabilized K\"ahler moduli.  Then, the mirror 
interpretation of this result is the number of vacua which stabilize
the {\blue volume of the compact dimensions} at a given value.
Clearly large volume is highly unnatural in this construction,
but do {\bf any} vacua reach the $V \sim 10^{30}$ of the ``large extra
dimensions'' scenario \cite{Antoniadis:1998ig} ? 

Writing $V\sim R^6$, we find a number of vacua
$$
N \sim {(2\pi L)^K R^{-K} \over K!}
$$
so large $K$ disfavors large volume in this case, and the maximum
volume one expects is of order 
$$
V \sim \left({2\pi L\over K}\right)^6 .
$$
where the parameter $L=\chi/24$ in F theory compactification on fourfolds.

In fact the maximal value we know of for $L$
is $L=75852$ \cite{KLRY}, but this comes with a large $K$ as well.
Now the effective $K$ might be reduced by imposing 
discrete symmetries, so a few large volume flux vacua might exist.
But the general conclusion is that large volume is highly disfavored within
this class of vacua.

\begin{figure}[t]
\vskip 5.5cm
\end{figure}

\section{Supersymmetry breaking}

Given a precise ensemble of effective field theories, 
such as the ensemble of \IIb\ theories on CY with flux \cite{GKP}
in which we assume that the effective potential is given by the
standard supergravity formula,
$$
V = e^K\left(g^{i\bar j} D_i W \bar D_{\bar j} W^* - 3|W|^2\right) + D^2 ,
$$
the problem of counting supersymmetry breaking vacua and finding their
distribution is a problem in mathematics, very similar to the problems
we just discussed of counting supersymmetric vacua.  We just want to
count critical points $V'=0$ with $V''$ a positive definite matrix
(so the vacua are metastable).  One of course needs to justify the
assumptions, but we return to this later.

D breaking vacua (with $DW=0$) are described by the earlier results, just
we require the vacua to be tachyon free and have near zero c.c.
Some results for F breaking flux vacua appear in \cite{DD}, and
we are continuing this study \cite{toapp}.  As we discussed, the
flux vacua which stabilize near conifold points are dual to the
hierarchically small scales arising from gauge theory,
so flux vacua should also include
the traditional scenario in which such effects drive susy breaking at 
small scales.
While our results are still preliminary, they are consistent
with the idea that this is a generic class of vacuum.

However, our results also describe another generic
class of vacuum not much discussed in previous literature, in which
the distribution of supersymmetry breaking scales is uniform
with a fairly high cutoff, possibly of order
$M_{str}^2$ and possibly lower (for example,
see eqns (4.31) and (4.32) in \cite{DD}).

While this remains to be verified in detail, the likely picture of the
distribution of supersymmetry
breaking scales in a one parameter model is the sum of uniform
and hierarchically small components, which we could model by the ansatz
$$
d\mu[M_{susy}] \sim c d(M^2_{susy}) + (1-c) d\log M_{susy} .
$$
with a parameter $c\sim 0.5-0.9$.
The intuition which leads to this is simply that there is nothing
inherent in the problem of supersymmetry breaking which favors low
or high scales (as long as $M_{susy}<<M_{Planck}$), so we should 
expect to see a distribution much like what we saw in
Fig. \ref{quinticsusydens} for the
scales which appear in supersymmetric vacua.

Now, once one believes in the existence of a significant population of vacua
with high scale supersymmetry breaking, it becomes conceivable that stringy
naturalness will not favor supersymmetry as a mechanism for solving the
hierarchy problem.  After all, the largest possible factor we could imagine
gaining through supersymmetry is about $10^{100}$ (from mitigating the
c.c. and hierarchy problems, as suggested in
\cite{mrd-stat,Banks,Douglas:2004kp}), 
and if the ratio of high scale to low scale vacua was larger
than this, then stringy naturalness would favor high scale breaking.

In fact, the factor by which supersymmetry improves the hierarchy
problem is much less than $10^{100}$, as argued in \cite{mrd-susy,%
Suss-susy,DGT}.  This argument is already subtle and interesting.
We can phrase it in terms of an estimate for the joint distribution of vacua
Eq. (\ref{eq:susydist}), which one might naively guess goes as
\begin{equation}\label{eq:wrong}
d\mu[M_{H}^2,M_{susy}^2,\Lambda] \sim
 \frac{dM_{H}^2}{M_{susy}^2}\ \frac{d\Lambda}{M_{susy}^4} d\mu[M_{susy}^2]
 \qquad (?)
\end{equation}
leading to the $10^{100}$ ratio.
However, both explicit factors in this formula are incorrect.  

One way to see that the factor $d\Lambda/M_{susy}^4$ is not right,
is to realize that it is based on the intuition that $\Lambda\rightarrow 0$
as $M_{susy}\rightarrow 0$, but in supergravity this is of course not true.
Rather, one needs to know 
the distribution of the parameter 
$3 e^K|W|^2$
which tunes away the vacuum energy from supersymmetry breaking,
\begin{equation}\label{eq:vtotal}
V  = M_{susy}^4  - 3 e^K|W|^2
= \sum_i |F_i|^2 + \sum_\alpha D_\alpha^2 - 3 e^K|W|^2
\end{equation}
As we saw in fig. \ref{CCdistr},
in flux vacua the parameter $e^K|W|^2$ is fairly uniformly
distributed, from zero
all the way to many times the string scale
\cite{DD}.  This means
that an arbitrary supersymmetry breaking contribution to the vacuum
energy, even one near but below the string scale,
can be compensated by the negative term, with no preferred scale.

The physics behind this is that the superpotential $W$ is a sum of
contributions from the many sectors.  This includes supersymmetric hidden
sectors, so there is no reason $W$ should be correlated to the scale
of supersymmetry breaking, and {no reason} the cutoff on the $W$
distribution should be correlated to the scale of supersymmetry breaking.
Such a sum over randomly chosen complex numbers will tend to produce
a distribution $d^2W$, uniform out to the cutoff scale.  For fluxes this is
the string scale, and this is plausible for supersymmetric sectors more 
generally.  Finally, writing $W=e^{i\theta}|W|$, we have
\begin{equation}\label{eq:Wdist}
d^2W = \frac{1}{2} d\theta d(|W|^2) = \frac{1}{2} d\theta d(|W|^2)
\end{equation}
leading very generally to the uniform distribution for $3|W|^2$.
Thus, the corrected version of Eq. (\ref{eq:wrong}) goes as
$d\Lambda/M_{str}^4$, and the need to get small c.c. {\bf does not
favor a particular scale of susy breaking} in these models \cite{mrd-susy}.

Next, the idea that a fraction $M_H^2/M_{susy}^2$ of models will
fine tune the Higgs mass is very likely incorrect as well, as pointed
out by Dine, Gorbatov and Thomas \cite{DGT}.  This is because
soft masses in models with a high scale of supersymmetry breaking
are naturally of the order $M_{3/2}^2=M_{susy}^4/M_{planck}^2$ (through 
non-renormalizable operators; gravitino loop effects and so forth).
A plausible summary
of the current understanding of this distribution is as follows:
\begin{itemize}
\item If the model contains no mechanism for solving the $\mu$ problem,
supersymmetry does not help at all (there is a 
supersymmetric mass term $\mu H_1 H_2$), and
we expect a fraction $M_H^2/M_{st}^2$ of models to realize the observed
Higgs mass.
\item If $M_{susy} > (M_H M_{pl})^{1/2} \sim 10^{10} \GeV$, then
$$
d\mu[M_H^2,M_{susy}^2] \sim \frac{M_H^2}{M_{susy}^4/M_{pl}^2}
\qquad\qquad M_{susy}^2 > M_H M_{pl}
$$
in a relatively model independent way.
\item
If $M_{susy} \le (M_H M_{pl})^{1/2} \sim 10^{10} \GeV$, 
we are in the situation of ``gauge mediation,'' in which the
leading coupling of supersymmetry breaking to the soft masses is
model dependent.  While we would need information about the distribution of
matter theories to say anything precise 
about this, it is reasonable to assume that
$M_H$ is roughly independent of $M_{susy}$, and
$$
d\mu[M_H^2,M_{susy}^2] \sim \epsilon \sim 1
\qquad\qquad M_{susy}^2 \le M_H M_{pl} .
$$
\end{itemize}

While this is already a bit complicated, we can draw from it the
conclusion that if the number of vacua
grows faster with  supersymmetry breaking scale than $M_{susy}^2$,
$$
d\mu[M_{susy}^2] \sim M^{2\alpha}_{susy} d(M^2_{susy})
 \qquad\qquad {\rm with}\ \alpha>1,
$$
then {\bf high scale breaking} will be favored.
While there would be many further points to make precise, this would
start to be the gist of an argument
predicting that {\bf we would not see superpartners at LHC.}

What makes this observation particularly interesting is that there is
in fact a very simple mechanism which could lead to a power law
growth of the number of vacua with supersymmetry breaking scale
\cite{Suss-susy,mrd-susy}.  It is that the total supersymmetry
breaking scale (which enters $M_{3/2}$ for example) is the sum
of positive quantities (as in Eq. \ref{eq:vtotal}).  Not only is there 
no possibility of the cancellations which led to Eq. (\ref{eq:Wdist}),
one can easily imagine that such a sum could produce a rapidly growing
distribution.

For example, convolving uniform distributions for each of the
individual breaking parameters, $d^2F$ and $dD$,  gives
\begin{eqnarray*}
\rho(M_{susy}^2) = \int &\prod_{i=1}^{n_F} d^2F \ %
\prod_{\alpha=1}^{n_D} dD 
\ dM_{susy}^4\\
& \delta(M_{susy}^4 - \sum |F|^2 - \sum D^2) \\
&\ \ \sim (M_{susy}^2)^{2n_F+n_D-1} dM_{susy}^2
\end{eqnarray*}
Now the inequality $2n_F+n_D\ge 2$ is surely satisfied by almost all
string models; indeed we see that even the distribution $d^2F$ (in
obvious analogy to Eq. \ref{eq:Wdist}) would already be on the edge.

While all this oversimplifies the real distributions, I believe it
does make the point that very simple and natural assumptions --
specifically, the existence of many hidden sectors, and the idea that
supersymmetry breaking can receive independent contributions in each
-- might in principle lead to so many high scale models that high
scale supersymmetry breaking becomes the natural outcome of string/M
theory compactification.

As I mentioned, Denef and I hope to get more definite results for
the distribution of susy breaking scales in flux vacua before long.  
There are of course
many more issues to consider: it might be that physics we neglected also
puts a lower cutoff on the maximal flux for supersymmetric vacua, it
might be that the $\mu$ problem is hard to solve, there might be large
new classes of nonsupersymmetric vacua (as suggested in \cite{silverstein}), 
etc.

\section{Conclusions}

We have gone some distance in justifying and developing the
statistical approach to string compactification.

We have specific \IIb\
orientifold compactifications in which all K\"ahler moduli are
stabilized, and vacuum counting estimates which suggest that all
moduli can be stabilized.
So far, it appears that such vacua are not generic, but they are
not uncommon either: about a third of our sample of F theory models
with Fano threefold base should work.

We have explicit results for distributions of flux vacua of
many types: supersymmetric, non-supersymmetric, tachyon-free.  They
display a lot of structure, with suggestive phenomenological
implications:
\begin{itemize}
\item Large uniform components of the vacuum distribution.
\item Enhanced numbers of vacua near conifold points.
\item Correlations with the cosmological constant.
\item Falloff in numbers at large volume and large complex structure.
\end{itemize}

There are intuitive arguments for some of the most basic properties.
For example,
the $-3e^K|W|^2$ contribution to the supergravity potential
is uniformly distributed with a large (at least string scale) cutoff,
because of contributions  from supersymmetric hidden sectors.
Thus, the need to tune the c.c. does not much influence the final numbers.

We start to see the possibility of making real world predictions:
\begin{itemize}
\item Large extra dimensions are heavily disfavored with the present
stabilization mechanisms.
\item Hierarchically small scales (gauge theoretic or warp factor)
are relatively common.
\item Uniform distributions involving many susy breaking parameters,
favor high scales of supersymmetry breaking.
\item 
This may imply that the gravitino and even superpartner masses should be
high, thanks to
supersymmetry breaking in hidden sectors.  
\end{itemize}

While various assumptions entered into the arguments we gave, the only
essential ones are that
\begin{itemize}
\item Our present pictures of string compactification are representative
of the real world possibilities.
\item The absolute number of relevant string/M theory compactifications
is not too high.
\end{itemize}
With further work, all the other assumptions can be justified and/or corrected,
because they were simply shortcuts in the project of characterizing
the actual distribution of vacua.

Since interesting results already follow from general properties of
the theory, and we now have evidence that the detailed distribution of
string/M theory vacua has many simple properties, we are optimistic
that a reasonably convincing picture of supersymmetry breaking and
other predictions can be developed in time for Strings 2008 at CERN.

\vskip 0.2in
I particularly thank my collaborators Frederik Denef, Bogdan Florea,
Bernie Shiffman and Steve Zelditch, and Tom Banks, Michael Dine, 
Gordy Kane, Greg Moore, Shamit
Kachru, Savdeep Sethi, Steve Shenker, Eva Silverstein,
Lenny Susskind, Scott Thomas, Sandip Trivedi, and
James Wells for valuable discussions and communications.

This research was partially supported by DOE grant DE-FG02-96ER40959.

\end{document}